\documentclass[12pt,epsf,aps]{revtex4}

\def\prl#1#2#3{{ Phys. Rev. Lett.} {\bf #1}, #2 (#3)}

\def\pra#1#2#3{Phys. Rev. A {\bf #1}, #2 (#3)}
\def\prb#1#2#3{Phys. Rev. B {\bf #1}, #2 (#3)}
\def\pre#1#2#3{Phys. Rev. E {\bf #1}, #2 (#3)}

\def\rmp#1#2#3{Rev. Mod. Phys. {\bf #1}, #2 (#3)}

\def\beq{\begin{equation}}
\def\bc{\begin{center}}
\def\ec{\end{center}}
\def\eqn{\end{equation}}
\def\la{\langle}
\def\ra{\rangle}

\begin{document}

\title{Pattern Formation in the Inhomogeneous Cooling State of Granular 
Fluids}

\author{Subir K. Das and Sanjay Puri} 
\address{School of Physical Sciences, Jawaharlal Nehru University \\
New Delhi -- 110067, India. \\}
\vspace{4cm}
\begin{abstract}

We present results from comprehensive event-driven (ED) simulations
of nonlinear pattern formation in freely-evolving granular gases. In
particular, we focus on the the morphologies of density and
velocity fields in the inhomogeneous cooling state (ICS). We emphasize
the strong analogy between the ICS morphologies and pattern formation
in phase ordering systems with a globally conserved order parameter.

\end{abstract}

\maketitle

\newpage

There has been much recent interest in the properties
of powders or granular materials \cite{jae,pg}. These materials are
of great scientific and technological relevance and display properties
intermediate to those of solids and fluids. Perhaps the most relevant
feature of powders is that the grains dissipate energy on
collision. Therefore, granular materials exhibit interesting dynamical
properties only when the collisional energy loss is compensated by a
continuous input of energy \cite{jae,pg}. 

It is also relevant to investigate the dynamical evolution
of a freely-evolving homogeneous system of inelastic granular particles.
In the initial stages, the system continuously loses
energy in a homogeneous cooling state (HCS), where the density field is
approximately uniform \cite{haf}. However, the HCS is unstable to density
fluctuations, and the system evolves into an inhomogeneous cooling state
(ICS), where particle-rich clusters are formed and grow \cite{gol}. There
is a good understanding of the HCS and the instabilities which result
in the ICS \cite{haf,gol,mcn,noi}. However, there is only a 
limited understanding of the nonlinear evolution of the ICS \cite{lh,bal}. 
In this letter, we study nonlinear domain growth processes for 
the granular density and velocity fields
in the asymptotic time-regime. We present results from
comprehensive event-driven (ED) simulations for a wide range of
inelasticity and density parameters.
In particular, we invoke analogies from studies of phase ordering 
dynamics \cite{op,bin} to obtain a {\it quantitative}
characterization of the evolving morphologies in the ICS.
The similarities between freely-evolving granular gases and phase
ordering systems have been briefly discussed by Van Noije,
Ernst and co-authors \cite{noi}, and Baldassari {\it et al.} 
\cite{bal}. Of course, we stress that the mechanisms driving pattern 
evolution are very different in both cases.

We consider a system of identical inelastic hard spheres
(with mass $m=1$ and diameter $\sigma=1$)
in a $d=2$ box of size $(N_b \sigma)^2$,
with periodic boundary conditions. There is no external input 
of energy in our simulations. We use an event-driven algorithm
\cite{all}, which only keeps track of particle collisions,
to evolve the system. After a collision between the 
$i^{th}$ and $j^{th}$ particles,
having velocities $\vec v_i$ and $\vec v_j$ respectively, the new
velocities are $\vec v_i^{\prime} = \vec v_i-\frac{1+e}{2}
[{\hat n}\cdot (\vec v_i-\vec v_j)]{\hat n},~~
\vec v_j^{\prime} = \vec v_j+\frac{1+e}{2}
[{\hat n}\cdot (\vec v_i-\vec v_j)]{\hat n}$.
Here, ${\hat n}$ is the unit vector parallel to 
the relative position of the particles.
For elastic collisions, we have the coefficient
of restitution $e=1$. For granular materials, $e<1$ in general. 

The initial condition for each run consisted of $N$ particles
with a homogeneous spatial distribution and a Maxwellian velocity
distribution. For circular particles, the average packing fraction
is $\phi=\pi N/(4N_b^2)$. We always fix $N_b=256$ and vary $N$, 
as specified below. We will label our results using the
number fraction, $n=N/N_b^2$. Numerical data are obtained 
for the following parameter sets:
(i) $N=30000$ (or $n\simeq 0.46, \phi\simeq 0.36$), and $e$ ranging from
0.80 to 0.975 in steps of 0.025;
(ii) $e=0.90$, and $N$ ranging from 10000 to 40000 in steps of 5000,
i.e., $n \simeq 0.15$ to 0.61. We characterize the dynamical 
evolution of the granular gas using various statistical quantities, 
which are calculated as an average over 5 independent runs, i.e., 
correlation functions, structure factors, and domain growth laws 
for the density and velocity fields.

For early times, inelastic collisions result in cooling as 
$\frac{dT(t)}{dt} = -\frac{\epsilon}{d} \omega (T) T$, 
where $T(t)$ is the temperature at time $t$; $\epsilon=1-e^2$;
and $\omega (T)$ is the collision frequency 
at temperature $T$ \cite{haf}. The form of $\omega (T)$
is approximately determined from the Enskog theory for elastic hard spheres
as $\omega (T) \simeq \omega (T_0) (T/T_0)^{1/2}$, 
where $T_0$ is the granular temperature at $t=0$.
This yields Haff's cooling law for the HCS \cite{haf} as
$T(t) = T_0 \left[1+\frac{\epsilon\omega(T_0)}{2d}t\right]^{-2}
= T_0 e^{-\frac{\epsilon}{d} \tau}$,
where $\tau (t)$ is the average number of collisions per particle upto time
$t$. Our subsequent results are presented in terms of $\tau$, as this
constitutes a reasonable measure of time in the present context.
The values of $\tau$ are obtained directly from the ED simulations.
We should stress that $\tau (t)$ vs. $t$
fluctuates considerably for individual runs in the ICS regime.

In Fig. 1, we show evolution snapshots
for the $d=2$ granular gas with $n \simeq 0.46$.
The LHS and RHS frames refer to the coarse-grained velocity field 
$\vec v (\vec r, \tau)$ and density field $\psi (\vec r, \tau)$,
respectively. The coarse-grained fields at a
lattice point are obtained by averaging over boxes of size $(5\sigma)^2$
centered at that point. To clarify the nature of
pattern formation, we have ``hardened'' the velocity field in
Fig. 1, i.e., the length of all vectors has been set to unity. Points
where the velocity field is zero, due to the absence of particles in
the coarse-graining box, are unmarked in Fig. 1. The density field
on the RHS of Fig. 1 is depicted in a binary representation.
We introduce the order parameter $\psi(\vec r,\tau)$ with values
$+1~(-1)$ at points where the number density is larger (less) than
the average number density.

The upper frames in Fig. 1 correspond to $e=0.85$ and $\tau = 150$. The
velocity field is characterized by the emergence and diffusive coarsening
of vortices. There is a progressive parallelization of the
local velocity field due to inelastic dissipation of the normal
velocity components. However, as the total momentum is fixed
at zero and the system is doubly-periodic, defects will be present
in the velocity field for all times. Typically, pattern
evolution in the density field is slower than that for the
velocity field \cite{noi}. Nevertheless, well-defined clusters 
corresponding to the ICS are already seen at $\tau=50$ (not shown here).
These clusters grow with time in a manner reminiscent of
phase ordering systems \cite{op}. The vortex centers
in the velocity field are also shown (as black circles) in the
frame on the RHS. The vortex density diminishes
with time, and vortices are primarily confined
to regions of low density, as there is a rapid parallelization of
velocities in the high-density (solid-like)
region due to multiple collision
processes. There is no strong correlation between the location of defects
in the velocity field (i.e., vortices) and defects in the
density field (i.e., interfaces or domain boundaries). The middle
frames in Fig. 1 correspond to $e=0.90$. 
The broad features are the same as those for $e=0.85$, though 
the time-scales are slower. All time-scales for pattern formation 
diverge as $e \rightarrow 1$, i.e., the clustering instability
is delayed for more elastic systems. The perfectly elastic case
($e=1$) is singular, and does not exhibit clustering.

A coarse-grained description of the granular fluid is provided
by nonlinear hydrodynamic equations for the density, velocity
and temperature fields, in conjunction with an energy loss term
\cite{gol}. Wakou {\it et al.} \cite{wak} have shown that
fluctuations in these quantities obey TDGL-like equations of
phase ordering dynamics with a nonconserved order parameter
\cite{bin}. Typically, these equations are obtained as 
the overdamped limit of a Hamiltonian formulation. In this
letter, we clarify the TDGL-description of granular dynamics.

The evolution of the density field is analogous to phase ordering dynamics
in two-component (AB) mixtures, which is described by the equation 
\cite{bin}:
\begin{equation}
\label{che}
\frac{\partial \psi(\vec r,t)}{\partial t}=
(-\nabla^2)^{m}\left[\psi(\vec r,t)-\psi(\vec r,t)^3+\nabla^2\psi(\vec r,t)
\right],
\end{equation}
where $\psi(\vec r,t)$ differentiates
between $A$-rich ($\psi=+1$) and $B$-rich ($\psi=-1$) regions.
Eq. (\ref{che}) with $m=0$ is the TDGL equation, and describes 
nonconserved systems. In this case,
the coarsening system is characterized by a diffusive 
growth law, $L(t)\sim t^{1/2}$ \cite{op,bin}, where $L(t)$
is the characteristic domain size at time $t$. On the other hand,
Eq. (\ref{che}) with $m=1$ is the Cahn-Hilliard (CH)
equation, and describes conserved systems.
In this case, the domain growth process obeys the
Lifshitz-Slyozov (LS) law, $L(t)\sim t^{1/3}$ \cite{op,bin}.
Finally, there have also been studies of Eq. (\ref{che}) with $m \rightarrow
0^+$, which is referred to as the globally-conserved (GC) TDGL equation
\cite{gc}. This model is known to be in the same dynamical universality
class as the nonconserved TDGL equation \cite{gc}, with the
difference that the initial composition is preserved in the
GC-TDGL equation.

Similarly, the evolution of the velocity field in the ICS 
is comparable to ordering dynamics in the XY model, which is 
the vector version of Eq. (\ref{che}):
\begin{equation}
\label{xym}
\frac{\partial \vec\psi(\vec r,t)}{\partial t}=
(-\nabla^2)^{m}\left[\vec\psi(\vec r,t)-|\vec\psi(\vec r,t)|^2
\vec\psi(\vec r,t) + \nabla^2\vec\psi(\vec r,t)\right],
\end{equation}
where $\vec\psi\equiv (\psi_1,\psi_2)$. The evolution in Eq. (\ref{xym})
parallelizes $\vec\psi$ locally via the annihilation of
vortices and anti-vortices, driven by a surface-tension reduction
mechanism. Eq. (\ref{xym}) with $m=0$ corresponds to the case with
nonconserved order parameter. The corresponding
domain growth law \cite{bp} is $L_v (t) \sim t^{1/2}$. On the other hand,
Eq. (\ref{xym}) with $m=1$ corresponds to the conserved XY model,
where the dynamics locally conserves $\vec \psi$.

The evolution of the granular fluid conserves both the density 
and velocity fields. Intuitively, one may expect that the 
analogous phase ordering models would be the conserved ($m=1$) 
versions of Eqs. (\ref{che})-(\ref{xym}).
However, the evolution morphologies in the upper and middle
frames of Fig. 1 are comparable to those for the case with
nonconserved order parameter \cite{op}.
This is a consequence of the non-diffusive dynamics of
granular particles, which move in straight lines until
they collide with other particles. In this collision, the density 
and momentum are conserved quantities, of course. However, depending
on the density of the fluid, the distance traveled by particles
prior to collision may be considerable. Therefore, the variables
are conserved on the macroscopic length-scale of the mean-free 
path, and not on the microscopic length-scale of the lattice spacing.
Now, the ICS consists of regions of high density and low density.
Granular particles stream relatively unhindered through the
low-density regions and deposit on distant clusters. Typically, the
conservation length scale is comparable to the length scale
of the coarsening clusters, which diverges in time.
Therefore, the density and velocity fields are globally
conserved (GC), rather than locally conserved. To elucidate this analogy,
the lower LHS frame in Fig. 1 shows the velocity field from
the GC-XY model with $\la \vec \psi \ra = 0$. The lower RHS frame in 
Fig. 1 shows the density field for the GC-TDGL model with $\la \psi \ra 
=-0.08$, corresponding to an average density of $0.46$.

Before we quantify the morphologies in Fig. 1, we would like to
study the HCS $\rightarrow$ ICS crossover time. 
Brito and Ernst \cite{be} use mode-coupling techniques to 
obtain the asymptotic energy decay (in the ICS) as
\begin{equation}
\label{mode}
T(\tau)\simeq \frac{T_0}{2n}\left(\frac{d-1}{\xi_{\perp}^{d}}
+\frac{1}{\xi_{||}^{d}}\right)
\left(\frac{4\pi\epsilon}{d}\tau\right)^{-d/2} ,
\end{equation}
where $\xi_{\perp} \simeq \sqrt{\frac{2d}{\epsilon}}l_0$ and
$\xi_{||} \simeq \frac{2d}{\epsilon} l_0$, with $l_0$
being the time-independent mean-free path.
A comparison of Haff's law and Eq. (\ref{mode}) yields the 
crossover time $\tau_c$ as the solution of
\begin{equation}
\label{tauc}
\tau_c^{d/2}e^{-\frac{\epsilon}{d}\tau_c}\simeq
\frac{(d-1)}{2}\left(\frac{\Omega_d}{4\pi}\right)^{d}
\chi(n)^{d}n^{d-1},
\end{equation}
where $\sigma = 1$, and $\Omega_d = 2\pi^{d/2}/\Gamma (d/2)$ is the
$d$-dimensional solid angle. We consider the case $\epsilon\rightarrow 0$,
so that $\xi_{||} \gg \xi_{\perp}$. 
For $d=2$, Eq. (\ref{tauc}) simplifies as
$\tau_c e^{-\frac{\epsilon}{2}\tau_c} \sim n\chi(n)^2$, where we
ignore prefactors. We use the Verlet-Levesque
approximation for the $d=2$ hard-sphere correlation function
at contact, viz., $\chi (n)=(1-7\phi/16)(1-\phi)^{-2}$.
In Fig. 2(a), we plot $e^{\frac{\epsilon}{2}\tau_c}$ vs. $\tau_c$
for $n\simeq 0.46$ and a range of $e$-values. 
In Fig. 2(b), we plot $\tau_c e^{-\frac{\epsilon}{2}\tau_c}$
vs. $n\chi(n)^2$ for $e=0.90$ and a range of $n$-values. 
These approximately linear plots confirm the validity of the 
scaling behavior of $\tau_c(e,n)$, as expected from Eq. (\ref{tauc}).

Next, we examine the time-dependent structure factors of the density and
velocity fields in the ICS. The evolving morphologies are 
characterized by unique length scales, and we expect the
structure factors to exhibit dynamical scaling, i.e., 
$S(k,\tau)= \ell (\tau)^{d} f(k \ell)$, where $\ell$ is the relevant
length scale and $f(x)$ is a scaling function \cite{bin}. In Fig. 3(a),
we plot the scaled structure factors of the density field for $(n,e)=(0.46,
0.85)$, i.e., $\ln [S_{\psi \psi} (k,\tau) \la k \ra^2]$ vs. 
$\ln (k/\la k \ra)$ from three different times ($\tau \gg \tau _c$). 
The quantity $\la k \ra$
is the first moment of the structure factor, and is related to the
length scale as $\la k \ra^{-1} \sim L (\tau)$.
The reasonable data collapse confirms the validity of dynamical
scaling. The dashed line in Fig. 3(a) denotes the Fourier transform 
of the analytic result due to Ohta {\it et al.} (OJK) \cite{ojk} for the 
correlation function of the nonconserved TDGL equation:
$g(x) = \frac{2}{\pi} \sin^{-1}(\gamma)$, where $\gamma=e^{-x^{2}}, x=r/L$. 
The OJK function provides an excellent description of our numerical data. 
The dot-dashed line in Fig. 3(a) is obtained from
a numerical simulation of the CH equation with $\la \psi \ra = -0.08$. 
The CH result is at variance with our numerical results except
for the tail region. In particular, the local conservation law dictates
that $S_{\mbox{\scriptsize CH}}(k,t) \sim k^4$ as 
$k \rightarrow 0$ \cite{bin}.
However, the ICS structure factor appears to decay almost
monotonically from $k=0^+$, as is usual for structure
factors in ordering problems characterized by a nonconserved order
parameter. (Of course, global density conservation dictates that 
$S_{\psi\psi} (0,\tau)=0$.) Fig. 3(b) is analogous to 3(a), but
corresponds to the case with $e = 0.90$.

We have confirmed numerically that the scaled 
structure factor for the density field
in the ICS has only a weak dependence on the off-criticality
$\langle \psi \rangle$, which measures the average density.
Furthermore, it is numerically comparable to the OJK
function for a broad range of $\langle \psi \rangle$-values.
The nature of defects in the ordering field dictates 
some general properties of the structure factor. For 
example, scattering off interfaces in the density
field gives rise to a power-law (or Porod) decay in the structure
factor tail, $S_{\psi \psi} (k,\tau) \sim k^{-(d+1)}$ for large $k$ 
\cite{por}, which is seen clearly in Figs. 3(a)-(b).

In Figs. 4(a)-(b), we plot scaled structure factors,
$\ln [S_{vv} (k,\tau) \la k \ra^2]$ vs. $\ln (k/\la k \ra)$,
for the velocity field in the ICS. The dashed line denotes the Fourier
transform of the Bray-Puri-Toyoki (BPT) function for ordering in the 
nonconserved XY model \cite{bp}:
\begin{equation}
\label{bpt}
h(x)=\frac{n\gamma}{2\pi}
\left[B\left(\frac{n+1}{2},\frac{1}{2}\right)\right]^2
F\left(\frac{1}{2},\frac{1}{2};\frac{n+2}{2};\gamma^2\right),
\end{equation}
where $\gamma= e^{-x^2}, x = r/L_v$; $B(x,y)$ is the beta function; and
$F(a,b;c;z)$ is the hypergeometric function.
The BPT result corresponds to defects with $O(n)$
symmetry, and the case with $n=2$ is relevant here. Again, we see that
our numerical data is described well by the BPT function, and is completely
different from the result for the conserved XY model (obtained
numerically, and denoted as a dot-dashed line in Figs. 4(a)-(b)).

We have seen that vortex defects characterize the morphology of the 
ICS velocity field. These vortices give rise to a power-law
decay for the structure factor tail, 
$S_{vv} (k,\tau) \sim k^{-(d+2)}$ for large $k$. The general BPT 
result in Eq. (\ref{bpt}) exhibits
a power-law tail, $S(k,t) \sim k^{-(d+n)}$,
and is relevant for the ICS in granular systems with $d>2$.

For the time-regimes considered here, a description in terms of
two uncoupled order parameters is reasonable. A more complete 
description of the asymptotic ordering dynamics should
account for {\it holes} in the velocity field due to absence of
particles in some regions of space. This is 
provided by a model with spin-vacancy phase separation in conjunction
with XY-like ordering of the spin variable. In the current
context, the relevant model has coupled nonconserved (or globally
conserved) dynamics for the two ordering fields. At a later stage, we 
will discuss in detail the applicability of this model to the ICS.

Finally, we would like to briefly discuss domain growth laws in the ICS.
We should stress that late-stage dynamics is affected by inelastic
collapse, where a group of particles undergo an infinite number
of collisions in a finite time \cite{my}. The results discussed
here correspond to the regime prior to inelastic collapse.
Subsequent to the HCS $\rightarrow$ ICS crossover, our numerical results
(not shown here) are consistent with diffusive growth, $L(\tau),
L_v (\tau) \sim \tau^{1/2}$ \cite{noi}.
Furthermore, our numerical data is consistent with an
asymptotic power-law behavior, $\tau (t) \sim t^{2/3}$, for a wide
range of parameter values. Therefore, in real time, the length scales
behave as $L(t), L_v (t) \sim t^{1/3}$ \cite{lh}. The domain growth law 
for the ICS density field is the same
as the LS growth law for phase separation of binary mixtures.

In conclusion, we have undertaken comprehensive ED simulations of nonlinear
pattern formation in the density and velocity fields of an inelastic
granular gas. We find that there is a  close analogy between
ICS morphologies and phase ordering systems with
globally conserved order parameters. The nature of defects
in the ordering fields dictates general properties of the relevant
structure factors and correlation functions. Furthermore, our
numerical results for the structure factors of the ICS density and
velocity fields are described well by analytic results for
nonconserved phase ordering systems. Clearly, the general formalism
of phase ordering dynamics is of great utility in diverse problems of
pattern formation as many features of evolution morphologies are
determined by general principles, e.g., defect structures, conservation
laws, etc. \\
\ \\
\ \\
{\bf Acknowledgements}

SP is grateful to M.H. Ernst, I. Goldhirsch, H. Hayakawa,
S. Luding, U.M.-B. Marconi and H. Nakanishi
for helpful discussions and critical inputs on this problem.
The authors are grateful to I. Goldhirsch and S. Luding for
their critical reading of this manuscript.

\newpage
\begin{center}
{\bf Figure Captions}
\end{center}

\noindent {\bf Figure 1:} Evolution pictures for the coarse-grained 
density and velocity fields in the ICS. 
The upper RHS frame shows the density field at $\tau=150$ for
$(n,e)=(0.46, 0.85)$. Regions with density greater than the
average are marked in grey. The black circles denote the
vortex centers of the velocity field. The upper LHS frame shows the
direction vectors of the corresponding velocity field. For clarity,
we only show a 32$^2$ corner (as depicted in the RHS frame) of the
256$^2$ lattice. The middle frames are the density and velocity fields
at $\tau=150$ for $(n,e)=(0.46, 0.90)$. The lower RHS frame shows an
evolution picture (at $t=25$) for the GC-TDGL equation
with $\la \psi \ra =-0.08$, corresponding to average
density 0.46. The lower LHS frame shows an evolution picture (at $t=25$)
for the GC-XY model with $\la \vec \psi \ra =0$. \\
\ \\
{\bf Figure 2:} Dependence of $\tau_c$ on system parameters. 
The crossover time is obtained from energy-decay plots (not shown 
here) as the point of deviation from Haff's law. The error bars are
defined by the symbol sizes.
(a) Plot of $e^{\frac{\epsilon}{2}\tau_c}$
vs. $\tau_c$ for $n\simeq 0.46$; and $e$ ranging from $0.80$ to $0.95$.
The solid line denotes the best linear fit to the data.
(b) Plot of $\tau_c e^{-\frac{\epsilon}{2}\tau_c}$
vs. $n\chi(n)^2$ for $e=0.90$; and $n$ ranging from $0.08$ to $0.53$.\\
\ \\
{\bf Figure 3:} (a) Scaled structure factors for the 
coarse-grained density field for $(n,e)=(0.46, 0.85)$. Data is
plotted for $\tau (\gg \tau _c) = 100, 150, 250$.
(b) Analogous to (a), but for $(n,e)=(0.46, 0.90)$. \\
\ \\
{\bf Figure 4:} Analogous to Fig. 3, but for the coarse-grained
velocity field.

\end{document}